\documentclass[10pt,conference]{IEEEtran}
\usepackage{eurosym}
\usepackage{amssymb}
\usepackage{amsmath}
\usepackage{amsthm}
\usepackage{paralist}
\usepackage{graphicx}
\usepackage[font=footnotesize, justification=centering, labelsep=period]{caption}
\usepackage [noadjust]{cite}
\newtheorem{def_}{Definition}
\newtheorem{thm_}{Theorem}
\begin{document}
\title{\textbf{\Large The Typed Graph Model -- a Supermodel for Model Management and Data Integration}}
\author{
\IEEEauthorblockN{Fritz Laux}
\IEEEauthorblockA{Fakult\"{a}t Informatik\\
Reutlingen University\\
D-72762 Reutlingen, Germany\\
email: fritz.laux@fh-reutlingen.de}
}

\maketitle
\begin{abstract}
In recent years, the Graph Model has become increasingly popular, especially in the application domain of social networks. 
The model has been semantically augmented with properties and labels attached to the graph elements. 
It is difficult to ensure data quality for the properties and the data structure because the model does not need a schema. 
In this paper, we propose a schema bound Typed Graph Model with properties and labels. 
These enhancements improve not only data quality but also the quality of  graph analysis. 
The power of this model is provided by using hyper-nodes and hyper-edges, which allows to present data structures on different abstraction levels. 
We prove that the model is at least equivalent in expressive power to most popular data models. 
Therefore, it can be used as a supermodel for model management and data integration.
We illustrate by example the superiority of this model over the property graph data model of Hidders and other prevalent data models, namely the relational, object-oriented, XML model, and RDF Schema. 
\end{abstract}

\IEEEpeerreviewmaketitle

\section{Introduction}
\label{sec:intro}

The popularity of the Graph Model (GM) stems primarily from its application to social networks,  medicine, scientific literature analysis, drug analysis, power and telephone networks.
The flexibility of the GM contributes to its popularity, but its schema-less implementations are prone to data quality problems. This was pointed out in our DBKDA paper \cite{Laux} in which we introduced the Typed Graph Model (TGM) with schema support.
The present work extends our findings about the TGM and provides a proof of the expressive power and demonstrates its superiority over most popular data models.

\begin{figure}[]
\centering
\includegraphics[width=0.25\textwidth]{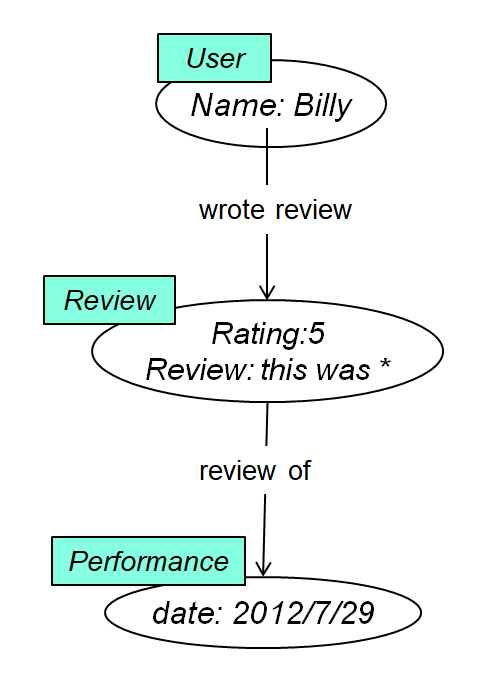}
\caption{Example graph taken partially from \cite{Robinson}, p. 42}
\label{fig:RobinsonsExample}
\end{figure}

Commercial graph database products like Neo4J \cite{neo4j}, ArangoDB \cite{arango}, JanusGraph \cite{janus}, Amazon Neptune \cite{neptune}, and others have been successfully applied to many domains. 
Advocates of the GM like Robinson et al. of Neo4J recommend in their book \cite{Robinson} to use specification by example, which builds on example objects. 
But this reaches not far enough as the following example taken from Robinson's book shows. 
It is depicted in Figure \ref{fig:RobinsonsExample} and shows a \emph{User} named Billy with its 5-star  \emph{Review} on a \emph{Performance} dated 2012/7/29. 
From this example we cannot know if Billy is allowed to have multiple reviews (on the same performance).
For good data quality, a review should depend on the existence of a user and a performance. 
But this cannot be derived from one example. 
This means that we have to deal with class things (like a generic Person) and not only with real objects (like Billy) and specify if a relationship is mandatory or optional.

In order to express structural information, it is necessary to abstract from a particular situation and specify integrity constraints. 
The use of a schema would help to ensure data integrity and would clarify the intended situation of the example.
Daniel et al. \cite{Daniel} complain that "there are only
few solutions that target conceptual modeling for NoSQL databases and even less
focusing on graph databases".  
They also point out the importance of a schema for data consistency and efficient implementation of a graph database and propose a framework, that translates an UML schema definition into a graph representation, and generate database-level queries from business rules and invariants.

Another weakness of the GM is that it has no notation to support different levels of detail and abstraction, which is apparently important for modeling large and complex data structures.

\subsection{Contribution}
To overcome these limitations we introduce in this paper a new typed graph model allowing hyper-nodes with complex structured properties (even sub-graphs) and hyper-edges connecting (recursively) one, two or more hyper-nodes. 
The graph schema provides data types, which allow type checking for instance elements. 
This ensures a formal data quality.
We prove that the model is at least equivalent in expressive power to most popular data models. 
It can be used on both, the instance and schema level. Its semantic power makes it suitable for a loss-less model management and high quality data integration.
Our model has a higher semantic expressiveness and precision than the prevalent data models, namely the relational, object oriented, XML data model, and RDF Schema. 
This will be demonstrated with typical modeling patterns.

\subsection{Structure of the Paper}
With the following overview of Related Work (Section \ref{sec:RelatedWork}) the context for our new typed graph model will be settled. 
Section \ref{sec:TypedGM} introduces and defines formally the Typed Graph Model (TGM) consisting of a typed schema and a hyper-graph instance connected to the schema. 
We present a compact and easy to read visualization of the model using UML.
The definitions are illustrated by some examples and the abstraction capability is demonstrated by the data model of a commercial enterprise.

In Section \ref{sec:Properties} we prove a semantic preserving schema translation for prevalent data models to the TGM. Properties of the TGM are explained and the translation process is illustrated for an Entity Relationship example.  
In the next Section \ref{sec:Comparison} our TGM is compared to the Graph Data Model (GDM) of J. Hidders \cite{Hidders}. 
Then, the semantic expressiveness of the TGM is demonstrated with typical data structures and
compared with the prevalent data models, namely the relational, object oriented, and XML data model.
The paper ends with a summary of our findings and gives an outlook on ideas for future work.

\section{Related Work}
\label{sec:RelatedWork}
Since the beginning of 1980 many papers on the GM have been published. DBLP \cite{dblp} alone retrieves 920 matches (retrieved July 30, 2021) for the key words "graph data model". If we ignore the papers that present specific applications for the GM incl. XML or Hypertext applications a few dozen of relevant papers remain. In the following, we discuss only papers that present the GM and its extensions (e. g., the Property Graph Model (PGM)) with a formal foundation or papers that use a graph schema:

The notion of PGM was informally introduced by Rodriguez and Neubauer \cite{Rodriguez}.
Spyratos and Sugibuchi \cite{Spyratos} use property graphs with hyper-nodes and hyper-edges for their graph data model. 
The main difference to our approach is that no schema is used and properties have no predefined data type. 
Another approach with hyper-edges is presented by Bu et al. \cite{Bu} who treats a label like a node connecting a set of nodes, which he calls hyper-edge. The nodes itself can be of different types. In this case Bu calls the graph a unified hyper-graph. The unified hyper-graph model is then applied to problems of ranking music content and combining it with social media information. Compared to our TGM the unified hyper-graph of Bu is only defined for graph instances. It is not not clear if the nodes have any type checking and if the whole graph is ruled by a schema. 

Ghrab et al. \cite{Ghrab} present GRAB, a schemaless graph database based on the PGM. It supports integrity constraints but cannot ensure data quality because of missing data types for properties and labels. Neo4J \cite{Robinson} has similar foundations and features. It has optional support for integrity constraints and comes with a powerful and easy to use graph query language, called \textit{Cypher}. 

All these PGM variants originate as instance graphs and no special attention is given to the graph schema.
No attempt is made to specify the different types of edges and the multiplicity of connections (edges) between different node types. 
Nodes are not typed and labels are not a proper substitute for data types. 
Therefore it is important to combine the PGM with a schema.

Amann and Scholl \cite{Amann} seem to be the first authors who connect a graph schema with its graph database instance. Nodes and edges do not have properties but both must conform to the schema. Their model is used for an algebra (hyperwalk algebra) for traversing the graph.

Marc Gyssens et al. \cite{Gyssens} and Jan Hidders \cite{Hidders} use a labeled GM to represent a database schema where each property of an object is modeled as a node in the graph. 
Labels are used to name node classes and edges.
The models become confusing because a node represents either an object, a property or a data type. 
Still, it is not possible to restrict the cardinality of schema edges (relationships).
Hidders' model is explained in more detail and compared to our TGM in Section \ref{sec:Comparison}.

Similar to Amann and Scholl the paper of Pab\'on et al. \cite{Pabon} uses a graph schema to query the graph database. 
They distinguish different node types, which they call "sort". 
The supported types are: \textit{object class nodes} (complex objects), \textit{composite-value class nodes} (for aggregate values), and \textit{basic-value class nodes} (primitive data types). 
This model seems to be equivalent to (complex) nodes with properties governed by a schema. 
A mechanism to abstract and group sub-graphs is missing, but would help to make the model easier to communicate.

Pokorn\'y \cite{Pokorny} uses a binary ER-Model as graph conceptual schema. 
For the graphical rendering he uses a compact entity representation for the nodes with attribute names inside the entity box. 
This solves the problem using the same node symbol for entities and attributes (properties) as it is the case with Gyssens \cite{Gyssens} and Hidders \cite{Hidders} models. 
The edge cardinality is represented in a form of crow-foot notation. 

In order to make the GM usable for real life scenarios with hundreds of schema elements, it is necessary to group or combine graph elements to higher abstracted objects. This would make the model easier to handle.

The need for grouping graph elements is addressed by Junghanns et al. \cite{Junghanns}. 
Their model allows to form logical sub-graphs (graph collections) with heterogeneous nodes and edges. 
With this it is possible to aggregate sub-graphs, e. g., user communities. 
The authors use UML-like graphical rendering of nodes to make the model better readable but their model fails to specify the cardinality of schema edges.

A step toward to complex composite nodes as an alternative approach to aggregation presents Levene \cite{Levene} by allowing the graph vertices to be recursively defined as a finite set of graphs. These hyper-nodes do not form a well-founded set as a node may contain itself, which violates the foundation axiom for the Zermelo-Fraenkel set theory. 

 A relatively new formal definition including integrity constraints was given by Angles \cite{Angles}. However, his model does not allow structured objects and grouping or aggregation. In the following section, we simplify his definitions and use it as basis for our TGM. 
 
\subsection{Comparison with Ontology Languages}
Ontology languages like the Resource Description Framework Schema (RDFS) \cite{RDFS} and the Web Object Language (OWL) \cite{OWL} are designed to specify ontologies and have their strength in allowing reasoning over instances of it. They are often used to semantically describe Linked Open Data (LOD) and the statement triples are usually visualized as graph structures.
RDFS and OWL provide a general type system that could be used to form user defined types. This would allow to use it as basis for a graph schema language. But if we look at the W3C OWL 2 Structural Specification \cite{OWLstructSpec} it seems difficult to define user specific classes and W3C itself uses UML class diagrams to illustrate OWL structures. 

Most approaches that map RDF to property graphs only support instance graphs. 
This is the case for the papers of Chiba et al., Sch\"atzle et al., and Nguyen et al. 
The paper of Chiba et al. \cite{Chiba} uses G2GML, a graph-to-graph mapping language where the source graph is a RDF-graph and the destination graph is a Property Graph (PG). 
With G2GML RDF patterns are specified in SPARQL syntax and the corresponding patterns of the PG are described in openCypher \cite{openCypher}. 
These patterns are mapped through \emph{node maps} and \emph{edge maps}. 
The main benefit is that the pattern specifications are domain-specific and declarative. There is no support for schema mappings.

Sch\"atzle et al. \cite{Schaetzle} define a mapping from RDF to the property graph model of GraphX. Their aim is to provide better analysis performance with S2X, a SPARQL implementation on top of GraphX and Spark, despite the schema-free PGM. Again, no schema mapping is supported.

The paper of Nguyen et al. \cite{Nguyen} represents each RDF triple element is  as a separate node, which justifies the name Labeled Directed Multigraph with Triple Nodes (LDM-3N). 
While other models represent predicates as labeled arcs, Nguyen et al. map them to nodes.
Assertions about RDF statements (triples) are modeled with the singleton property and not by reification. 
This approach adds an extra computation step and doubles the number of triples, which bloats the graph model. 
The main application domain seems to be the analysis of RDF triples by mapping it to the LDM-3N graph model allowing the use of graph analysis algorithms.
Finally, it does not support schema mapping.

The specification of data structures is not the core intention of RDF. In RDFS for instance it is not possible to define the cardinality of relationships. Likewise, OWL Lite has strong limitations on allowing only 0 or 1 as multiplicity of properties. Simple unique requirements and relations like one-to-one, one-to-many and many-to-one are cumbersome to define even in OWL Full.
Complex data structures need a modeling language that allows to define different levels of abstraction, which is not the strength of these ontology languages. Most examples of RDFS or OWL do not care about the multiplicity of relationships (cardinalities may be guessed via property names) and grouping of attributes seem to be on the same level as objects or subjects. 

All these arguments and examples make it clear that we need schema support to ensure high data quality when using graph databases. This can be achieved with the Typed Graph Model (TGM), which we develop in the next section.

\section{The Typed Graph Model}
\label{sec:TypedGM}

Our TGM informally constitutes a directed property hyper-graph that conforms to a schema.
In the following definitions our notation uses small letters for elements (nodes, edges, data types, etc.) and capital letters for sets of elements. Sets of sets are printed as bold capital letters. A typical example
would be $n \in N \in \mathbf{N} \subseteq \wp(N)$, where $N$ is any set and $\wp(N)$ is the power-set of $N$.

\subsection{Graph Schema}
\label{ssec:Schema}
 
Let $T$ denote a set of simple or structured (complex) data types. 
A data type $t := (l,d) \in T$ has a name $l$ and a definition $d$. 
Examples of simple (predefined) types are $(int, \mathbb{Z})$, $(char, ASCII)$, $(\%,[0..100])$ etc.
It is also possible to define complex data types like an order line $(OrderLine, (posNo, partNo,  partDescription, quantity))$. 
The components need to be defined in $T$ as well, e. g., $(posNo, int > 0)$. 
Recursion is allowed as long as the defined structure has a finite number of components.  

\begin{def_}[Typed Graph Schema]
A typed graph schema is a tuple $TGS = (N_S, E_S, \rho, T, \tau, C)$ where:
\begin{itemize}
\item
$N_S$ is the set of named (labeled) objects (nodes) $n$ with properties of data type $t := (l,d) \in T$, where $l$ is the label and $d$ the data type definition.
\item
$E_S$ is the set of named (labeled) edges $e$ with a structured property $p := (l,d)\in T$, where $l$ is the label and $d$ the data type definition.
\item
$\rho$ is a function that associates each edge $e$ to a pair of object sets $(O,A)$, i. e., $\rho(e) := (O_e,A_e)$ with $O_e,A_e \in \wp(N_S)$. $O_e$ is called the \emph{tail} and $A_e$ is called the \emph{head} of an edge $e$.
\item
$\tau$ is a function that assigns for each node $n$ of an edge $e$ a pair of positive integers $(i_n,k_n)$, i. e., $\tau_e(n) := (i_n, k_n)$ with $i_n \in \mathbb{N}_0$ and $k_n \in \mathbb{N}$. 
The function $\tau$ defines the min-max multiplicity of an edge connection. 
If the min-value $i_n$ is $0$ then the connection is optional.
\item
$C$ is a set of integrity constraints, which the graph database must obey. The constraint language may be freely chosen.
\end{itemize}
\end{def_}

The notation for defining data types T, which are used for node types $N_S$ and edge types $E_S$, can be freely chosen. 
The integrity constraints $C$ restrict the model beyond the structural limitations of the multiplicity $\tau$ of edge connections. 
Typical constraints af $C$ are semantic restrictions of the content of an instance graph.
This makes the expressiveness of the TGS at least as strong as the models to which it is compared in Section \ref{sec:Comparison}. 

\subsection{Typed Graph Model}
\label{ssec:TGM}

\begin{def_}[Typed Graph Model]
A typed graph Model is a tuple $TGM = (N, E, TGS, \phi)$ where:
\begin{itemize}
\item
$N$ is the set of named (labeled) nodes $n$ with data types from $N_S$ of schema TGS.
\item
$E$ is the set of named (labeled) edges $e$ with properties of types from $E_S$ of schema TGS.
\item
$TGS$ is a typed graph schema as defined in Subsection \ref{ssec:Schema}.
\item
$\phi$ is a homomorphism that maps each node $n$ and edge $e$ of $TGM$ to the corresponding type element of $TGS$, formally:
\begin{eqnarray*}
 \phi: & TGM &\rightarrow TGS \\
 & n &\mapsto  \phi(n) := n_S (\in N_S)\\
 & e &\mapsto  \phi(e) := e_S (\in E_S)
\end{eqnarray*}
\end{itemize}
\end{def_}
The fact that $\phi$ maps each element (node or edge) to exactly one data type implies that each element of the graph model has a well defined data type. 
The homomorphism is structure preserving. This means that the cardinality of the edge types are enforced, too. Data type and constraint checking is applied for all nodes and edges before any insert, update, or delete action can be committed. If no single type can be defined, union type or \emph{anyType} (sometimes called \emph{variant}) may be applied. Usually this is an indication for a weak data model and it should be clear that this could affect data quality and processing. 

As graphical representation for the TGS we adopt the UML-notation for nodes and include the properties as attributes including their data types. 
Labels are written in the top compartment of the UML-class.
Edges of the TGS are represented by UML associations.
For the label and properties of an edge we use the UML-association class, which has the same rendering as an ordinary class but its existence depends on an association (edge), which is indicated by a dotted line from the association class to the edge. 
This not only allows to label an edge but to define user defined edge types.

The correspondence between the UML notation and the TGS definition is is shown in Table \ref{TGSvsUML}.\\

\begin{table} [ht]
\caption{TGS correspondence with UML notation}\label{TGSvsUML}
\centering
\begin{tabular}{l | l}
TGS & UML \\
\hline
$n \in N_S$ & class \\
$e \in E_S$ & association \\
$t=(l,d) \in T$ & $l$ = name of $n$ resp. $e$; $d$ = type of $n$ resp. $e$ \\
$\rho(e)$ & all ends of $e$ \\
$\tau_e(n)$ & (min,max)-cardinality of e at n \\
$C$ & constraints in [ ] or \{ \} \\
\end{tabular}
\end{table}

The use of hyper-nodes $n \in N_S$ and hyper-edges $e \in E_S$ instead of simple nodes resp. edges allow to group nodes and edges to higher abstracted complex model aggregates. This is particularly useful to keep large models clearly represented and manageable.
Large graph models may then be grouped into sub-graphs like in Junghanns et al.\cite{Junghanns}. 
Each sub-graph can be rendered as a hyper-node. 
If the division is disjoint these hyper-nodes are connected via hyper-edges forming a higher abstraction level schema (see Figure \ref{fig:Enterprise} (b)).

\subsection{Examples}
\label{ssec:Examples}
\begin{figure}[]
\centering
\includegraphics[width=0.48\textwidth]{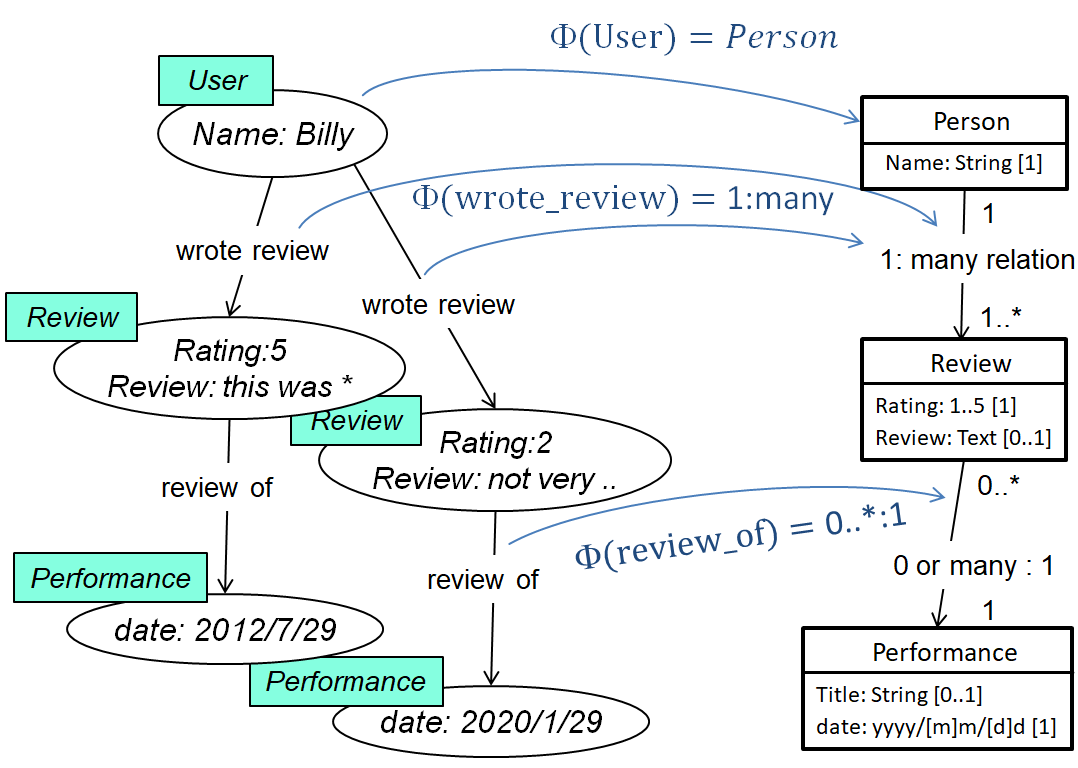}
\caption{Example graph with schema in UML notation}
\label{fig:SchemaRobinsons}
\end{figure}

Lets recall the example graph from Figure \ref{fig:RobinsonsExample} and model its corresponding schema. 
We want to make clear that a user may write as many reviews as he likes, but only one for a particular performance. 
A rating needs to refer exactly to one performance and one user. This is reflected in Figure \ref{fig:SchemaRobinsons} by the "1:many" and "0 or many:1" relationships.
We use the UML-notation for the schema and keep the notation from Figure \ref{fig:RobinsonsExample} for the instance graph for clarity.

The homomorphic mapping $\phi$ guaranties that the instance graph obeys the schema, i. e. type, cardinality, and constraint checking.
Now, it is clear from the schema that a user must have at least one review. The review is existence dependent on the user and a performance. The "wrote review" edge is a 1:many relation and "review of" is an optional many:1 relation. This has the consequence that a review needs a person and a performance. But, a performance may exist without any review.

\begin{figure*}[]
\centering
\includegraphics[width=0.7\textwidth]{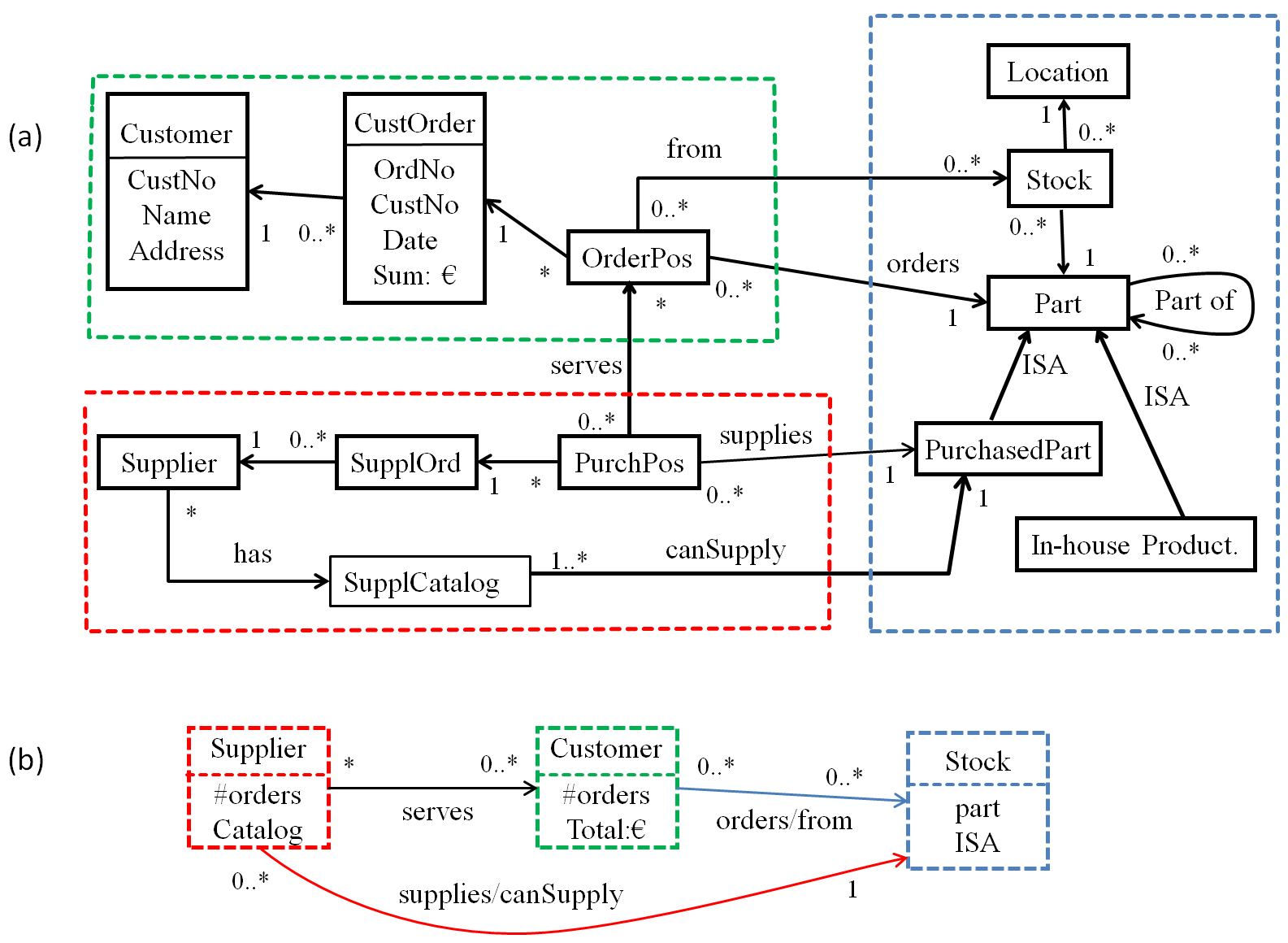}
\caption{Example TGM of a commercial enterprise showing two levels of detail}
\label{fig:Enterprise}
\end{figure*}

In the next example we present a commercial enterprise that sells products and parts to customers. 
The enterprise assembles products from parts and if the stock level is not sufficient it purchases parts from different suppliers.
Figure \ref{fig:Enterprise} models this situation using UML rendering. It demonstrates the abstraction power of the TGM showing two schema abstraction levels. 
The upper part (a) shows the TGM on a detailed level. 
The properties are suppressed in the diagram for simplicity except for \emph{Customer} and \emph{CustOrder}. The schema is grouped into 3 disjoint sub-graphs depicted with dashed lines. 

In the lower part (b) these sub-graphs are shown as hyper-nodes of the graph schema. 
This allows a simplified and more abstracted view of the model. 
Also, some aggregate properties (e. g. \#orders) are shown to illustrate the modeling capabilities.
The coloring of the (hyper-)edges helps the reader to identify which edges have been aggregated. 
The hyper-edges connecting these abstracted nodes must use the most general multiplicity of the multiple edges it combines. 
In the example the edge \emph{orders/from} combines two edges, i. e., \emph{orders} with 0..1 - 1 multiplicity and \emph{from} with 0..* - 0..* multiplicity, which leads to the most general multiplicity.

\section{Properties of the Typed Graph Model}
\label{sec:Properties}

The TGM has some valuable properties and can be regarded as a \textbf{supermodel} because its semantic expressiveness proves to be at least as powerful as prevalent data models. In order to prove this statement we need the following definitions. 

As we are only interested in databases that satisfy the constraints of its schema, we modify the definitions from Angles et al. \cite{Angles2020} and restrict it to schema bound databases. 
More precisely, let $M$ be a database model and $S^M$ be a database schema. A database is then an ordered pair $D^M := (S^M, I^M)$, where $I^M$ is an instance obeying all constraints of $S^M$.

\begin{def_}[Schema Mapping]
\label{def:SchemaMapping}
Let $M_1$ and $M_2$ be two data models. 
A schema mapping from $M_1$ to $M_2$ is a total function $\mathbf{SM}$ from the set of all database schemas in $M_1$, to the set of all database schemas in $M_2$. 
\begin{eqnarray*}
\mathbf{SM}: M_1 \rightarrow  M_2 \\
s_1 \; \mapsto \; s_2 \,
\end{eqnarray*}
where $s_1$ and $s_2$ are schemas of $M_1$ resp. $M_2$.
\end{def_}

\subsection{Some Property Definitions}
\label{ssec:Property}
Every data model allows to structure the data according to its modeling elements. 
These conceptual elements determine the representational power of the data model. 
A model $M_2$ subsumes the information capacity of $M_1$ if and only if every schema in $M_1$ can be translated to a schema in $M_2$ without loss of information.
Two database models can be evaluated in terms of its information capacity considering the following properties:

\begin{def_}[Computable Mapping]
A schema mapping $CM$ from $s_1 \in M_1$ to $s_2 \in M_2$ is computable if there exists an algorithm that translates schema $s_1$into $s_2$.
\end{def_}
If any schema mapping from $M_1$ to $M_2$ is computable then this implies that $M_2$ subsumes $M_1$. 
A computable mapping may still result in a schema that allows an invalid database instance.

\begin{def_}[Semantics Preservation]
A computable schema mapping $SP$ is semantics preserving if for every valid database $d_1$ of schema $s_1$, there is a valid database $d_2$ obeying schema $s_2$ where $s_2$ is produced by the
mapping $SP$, i. e. $s_2 = SP(s_1)$.
\end{def_}
This property guarantees that the result of the instance mapping will always be a valid database according to $s_2$.

\begin{def_}[Information Preservation]
A semantics preserving schema mapping $IP$ is information preserving if there exists an inverse computable schema mapping $IP^{-1}$ from $M_2$ to $M_1$ such that for every $s_1 \in M_1$ it holds $s_1 = IP^{-1}(IP(s_1))$. 
Such a schema mapping $IP$ is alternatively called "schema translation".
\end{def_}
This definition indicates that, for some schema mapping $IP$, there exists an "inverse" mapping which allows recovering the original schema previously transformed. 
In general, the inverse mapping $IP^{-1}$ is only a partial function because it is only defined on $s_2$ and may not be defined on all elements of $M_2$, i. e. the image $SM(M_1)$ can be a proper subset of $M_2$. 
Information preservation is an important property because it guarantees that a schema mapping results in a new schema capable to not lose any information.  
Moreover, it implies that the target database model $M_2$ subsumes the information capacity of the source database model $M_1$. If $M_1$ subsumes $M_2$ as well, then both data models are equivalent and $IP^{-1}$ and $IP$ are total functions. 
McBrien and Poulovassilis \cite{Poulovassilis}\cite{McBrien} describe equivalence-preserving mapping of schema constructs based on a Hypergraph Data Model (HDM). The ideas and problems with schema translation have been reviewed and revised in the light of model management by Bernstein and Melnik \cite{Bernstein}.

\subsection{Schema Mapping by means of a Meta-model}
\label{ssec:Mappings}

When considering database mappings two types of mappings can be distinguished: (1) schema independent and (2) schema dependent. 
We are only interested in schema dependent mappings to always ensure high data fidelity. 
The above schema mapping of Definition \ref{def:SchemaMapping} is model independent \cite{Lenzerini}. 
A meta-model that is general enough to capture all popular models in the literature would suffice. 
Hull \cite{Hull} as well as Atzeni and Torlone \cite{Atzeni1995}\cite{Atzeni1996} describe such a framework for heterogeneous data models. It was developed towards a model independent \textbf{supermodel} \cite{AtzeniEDBT06} and later implemented as a general tool, called MIDST \cite{AtzeniSIGMOD07, AtzeniVLDB08, AtzeniEDBT09}. 
It consists of the following meta-constructs: \emph{lexical, abstract, aggregation, generalization, and function}. 
For example, the Entity-Relationship Model (ERM) involves (i) abstracts (the entities), (ii) aggregations of abstracts (relationships), and (iii) lexicals (attributes) with functions (to entities or relationships). 
This means that the ERM is a specialization of the supermodel, i. e. a schema in any (sub)model is also a schema in the supermodel, only the names of the model elements differ.
Hull and Atzeni give more examples and claim that this supermodel subsumes Relational Model (RM), ERM, XML, Object Oriented Model (OOM), Object Relational (OR), and XSD. 

\subsection{Information Preserving Schema Translation to TGM}
\label{ssec:Semantics}
There are works that show information preserving mappings from ERM to Graph Database Schema \cite{RoyHubara}, RM to RDF and OWL \cite{Sequeda}, and RM, CSV, XML JSON to RDF using RML, the RDF Mapping Language \cite{Dimou}. 
Angles et al. \cite{Angles2020} have shown that a RDF database can be mapped to a property graph database.
This includes information preservation for the schema translation and the instance mapping. 
Taking all together it seems possible to translate most popular data models to an enhanced graph data model, i. e. a PGM.
It is evident that the TGM subsumes the PGM as any model element of PGM can be mapped 1:1 to the corresponding TGM model element.
The more general approaches of Arenas et al. \cite{Arenas}, and McBrien/Poulovassilis \cite{McBrien} provide criteria for model independent schema mappings that are information preserving. Hull \cite{Hull} and Atzeni/Torlone \cite{Atzeni1995}\cite{Atzeni1996} propose basic meta-constructs for the supermodel that covers all relevant data models. This puts us now in a position to state our main Theorem.

\begin{thm_}[Information Preserving Schema Translation to TGM]
\label{thm:TransToTGM}
Let $M$ be any data model that can be subsumed by the \textbf{supermodel} of Hull and let $T$ be the Typed Graph Model. For any schema $s \in M$ there exists an information preserving mapping (translation) $\mathbf{M}$:
\begin{eqnarray*}
\mathbf{M}: M \rightarrow  T \\
s \; \mapsto \; t \,
\end{eqnarray*}
where $t$ is a TGS and $s$ is a schema from $M$.
\end{thm_}

It would be easy to proof the theorem by contradiction. 
But such a proofs gives no constructive idea how a concrete mapping would look like. 
Therefore, we proof the theorem by constructing a generic mapping $\mathbf{M}$ taking the meta-constructs of the supermodel and assign uniquely model elements from $T$. 
These model element pairs allow us to map any schema $s \in M$ to a schema $t \in T$.
Then it will be shown that we can construct an inverse mapping that leads to the original schema $s$ again.

\begin{proof}[Proof of the Translation Theorem]
Following Atzeni and Torlone we define the schema mapping $\mathbf{M}(M) = T$ by the following elementary 1:1 transformations for each model element $\sigma \in s$:
\begin{enumerate}
\item lexical $\to$ property
\item abstract $\to$ node 
\item aggregation $\to$ edge with aggregation type
\item generalization $\to$ edge with generalization type
\item function $\to$ edge with a single target (multiplicity 1)
\end{enumerate}
The above transformation can be chained or composed to form a directed acyclic graph where each transformation assignment represents a node and the directed edges represent the sequence of assignments. 
Each path from a leaf node ends at the root node representing the mapping $\mathbf{M}$. 
Given a schema $s$ the mapping $\mathbf{M}$ translates the input $s$ into a TGS $t \in T$. 
All transformation steps are 1:1 such that the resulting function $\mathbf{M}$ is a injective mapping that can be reversed. 
The inverse mapping $\mathbf{M^{-1}}$ uses just the opposite transformation assignments listed above. 
It should be noted that $\mathbf{M^{-1}}$ may be a partial function only, i. e. there might exist model elements in $T$ (e. g. a composition edge type with existential dependency of its components) that may have no corresponding model element in the source model $M$.

Let $\sigma$ w.l.o.g. be any model element from schema $s$. $\mathbf{M}$ translates $\sigma$ into an element $\tau \in T$ in the following way:
\begin{eqnarray*}
\tau = \mathbf{M}(\sigma) = \mathbf{m_n(m_{n-1}( ...(m_1}(\sigma) ...))
\end{eqnarray*}
where $\mathbf{m_i}$ are the elementary transformations from above.
The data type $\sigma_t$ of an element $\sigma$ is carried over to the same data type of the translated element $\tau$.
Applying the inverse transformations $\mathbf{m^{-1}_i}$ in opposite order results in the identiy function 
$\mathbf{I}$ which proofs the information preservation property.
\begin{eqnarray*}
\sigma = \mathbf{m^{-1}_1(...(m^{-1}_{n-1}(m^{-1}_n(m_n(m_{n-1}( ...(m_1}(\sigma) ...))\\
 = \mathbf{M^{-1}(M}(\sigma)) = \mathbf{I}(\sigma))
\end{eqnarray*}
\end{proof}

\subsection{Example Translation from extended ERM to TGM}
\label{ssec:ERMtoTGM}

\begin{figure}[]
\centering
\includegraphics[width=0.4\textwidth]{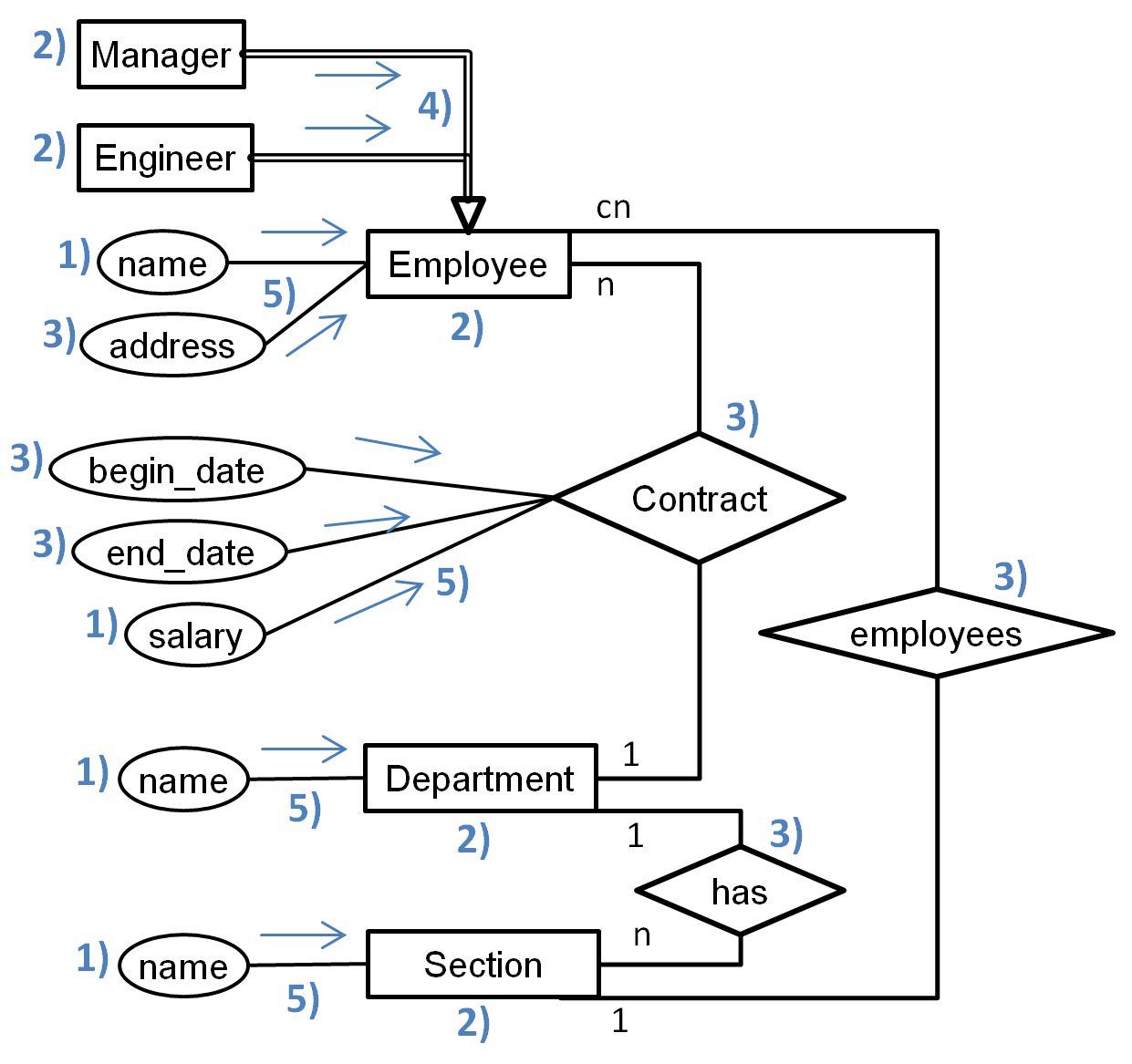}
\caption{Hidders' example \cite{Hidders} transformed to a TGS}
\label{fig:ERMtoTGM}
\end{figure}

To demonstrate the translation algorithm we use an example from Hidders' GDM \cite{Hidders}, which will be discussed and compared to the TGM in detail in the next section. 
At the moment, we concentrate on the translation process taking Hidders' example but using the ERM enhanced with generalization (IS-A) relationship. The original visualization of Hidders is depicted in Figure \ref{fig:CompareHidders} (a).
The extended Entity-Relationship (ER) diagram is shown in Figure \ref{fig:ERMtoTGM} where the blue numbers next to the ER-symbols refer to the elementary transformations from Theorem \ref{thm:TransToTGM}. 

The complete translation has to execute the elementary transformation steps from left to right (in order of the blue arcs) until all ERM elements have been converted. 
For instance, before the relationship \emph{Contract} can be mapped to an edge of general aggregate type as indicated by transformation 3), the relationship attributes need to be mapped first. 
The \emph{salary} is considered as a literal (lexical 1) that transforms directly to a \emph{Contract} property. The \emph{begin\_date} and \emph{end\_date} are considered as aggregation model elements because of its date structure. All three attributes are single valued (indicated by the function 5) and mapped as properties of \emph{Contract}.

It is also possible to specify more precisely the type of aggregation in the TGM by defining an edge type "Contract" that already includes the properties \emph{begin\_date}, \emph{end\_date}, and \emph{salary}.
This would result into the same semantics but with a user defined edge type.
It should be pointed out that this possibility results from TGM's capability to support different abstraction levels.

The result of the translation is shown in the lower part of Figure \ref{fig:CompareHidders}. 
The visualization of the TGS uses the UML rendering listed in Table \ref{TGSvsUML}.

\section{Comparison with other Data Models}
\label{sec:Comparison}

In the following, we compare our TGM to other models with respect to structural differences and schema support. We point out modeling restrictions of these models and show how such situations are modeled with TGM. Query and manipulation languages are beyond the scope of this paper.

\subsection{Comparison with GDM of Jan Hidders}
\label{ssec:Hidders}

Jan Hidders' \cite{Hidders} model added labels and properties together with their data types to nodes and edges  (relationships).
Property names are modeled as edges in the schema.
This allows to model labeled relationships 
with complex properties. 
Structured and base data types share the same graphical representation, which makes it difficult to distinguish both. 
The ISA-relationship is rendered as a double line arrow similar to the extended ERM. 
Hidders' model does not allow to restrict the cardinality of relationships. This restriction limits its modeling power compared to the TGM, which provides a min-max notation for the cardinality.

\begin{figure}[]
\centering
\includegraphics[width=0.45\textwidth]{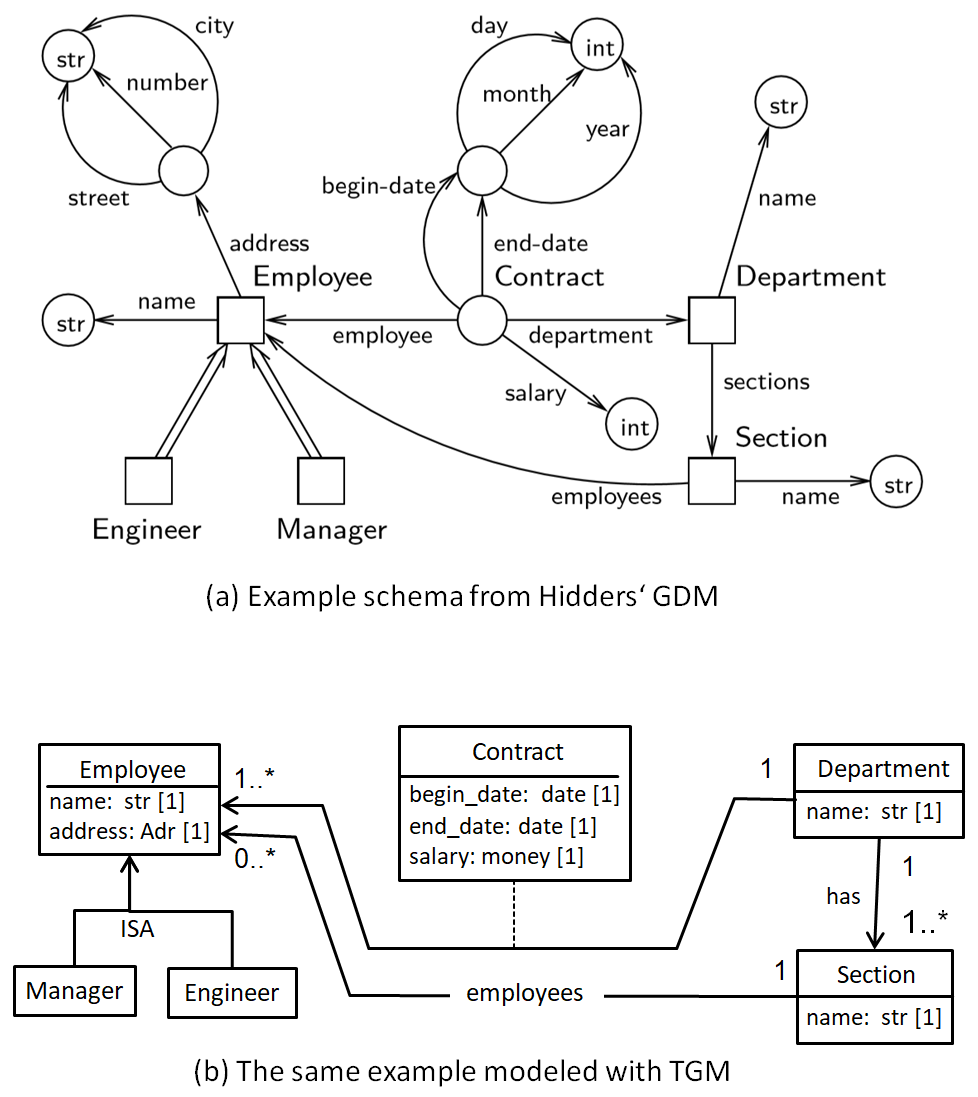}
\caption{Comparison by example with Hidders' GDM}
\label{fig:CompareHidders}
\end{figure}

The example in Figure \ref{fig:CompareHidders} is from the publication of Hidders \cite{Hidders}. 
The schema shows \emph{Employee} and \emph{Department} classes linked by a \emph{Contract}. The relationship \emph{Contract} is existence dependent on the connected nodes. The properties of \emph{Contract} are salary of type \emph{int}, begin-date and end-date of structure-type $date = (day, month, year)$.
In Hidders' model these dates are modeled on the element level using data type \emph{int}.
Hidders' schema elements, i. e., nodes (objects), edges (properties) and data types appear on the same visual level, which makes it difficult to read and obscures semantics. The modeling power of complex data types provide a clear advantage for the TGM.

\subsection{Comparison with the Relational Model (RM)}
\label{ssec:RM}

There is a 1:1 correspondence between attributes and properties and any relation can be modeled as a node  with properties. 
The min-max notation for relationship multiplicity can model any link cardinality.
The TGM can therefore easily represent tabular structures, foreign key constraints (many-to-one relationships), and join-tables as the building blocks of the RM. 
Beyond this, the TGM is able to directly model many-to-many relationships of any min-max multiplicity. This makes the TGM strictly stronger than the relational model.

\begin{figure}[]
\centering
\includegraphics[width=0.4\textwidth]{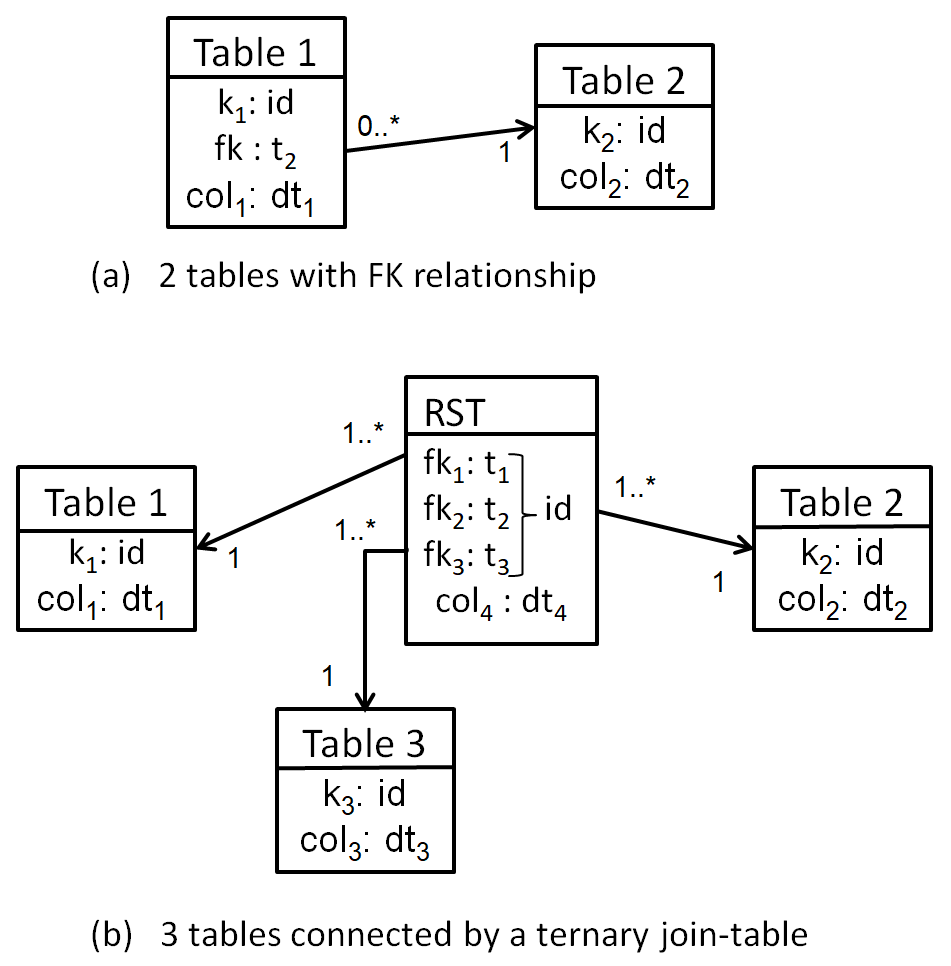}
\caption{Modeling a many-to-one relationship (FK) and a ternary join-table with the RM}
\label{fig:RM}
\end{figure}

Figure \ref{fig:RM} (a) shows a typical FK relationship between Table 1 and Table 2. In part (b) of Figure \ref{fig:RM} three tables are liked via a join-table RST with foreign keys $fk_1, fk_2, fk_3$ representing a many-to-many ternary relationship with attribute $col_4$.

\begin{figure}[]
\centering
\includegraphics[width=0.45\textwidth]{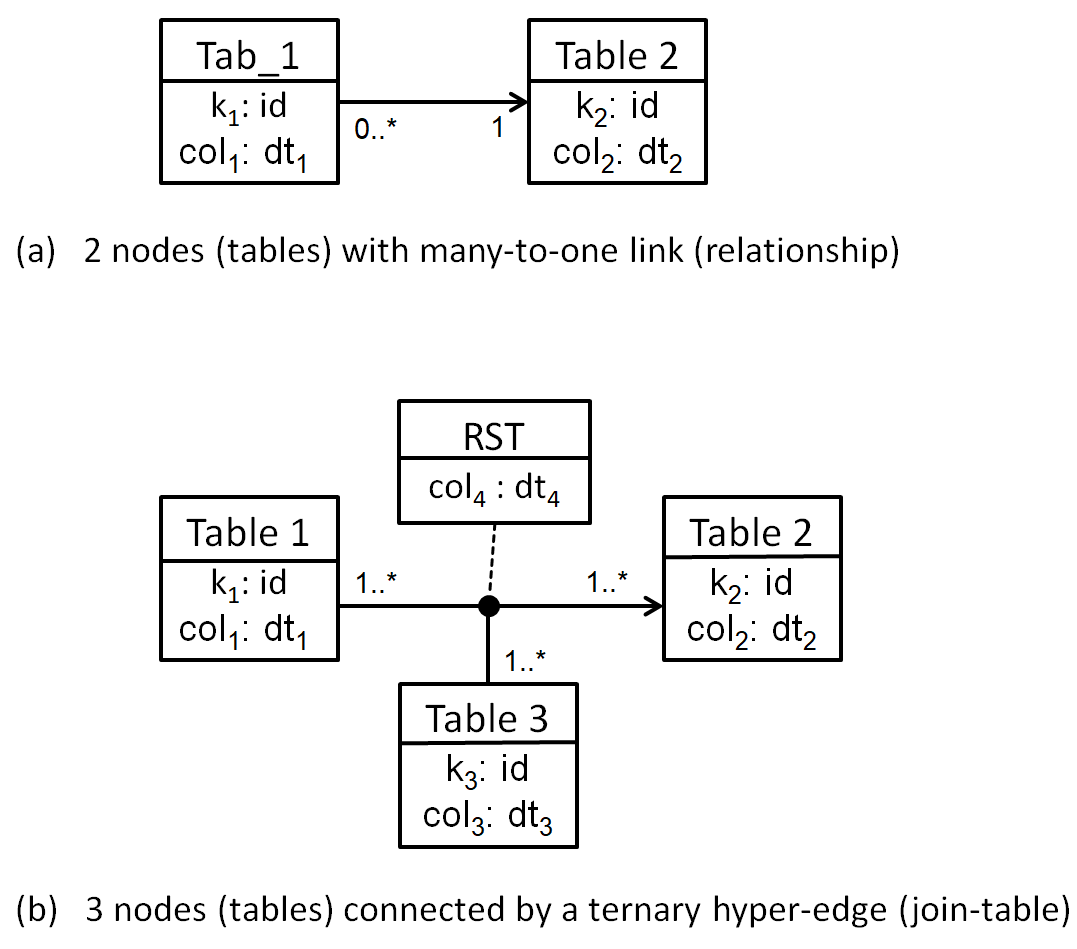}
\caption{Modeling a many-to-one relationship (FK) and a ternary join-table with TGM}
\label{fig:CompareRM}
\end{figure}

Another difference between TGM and RM is that foreign keys (FK) are not necessary because their function is taken over by an edge linking the Tab\_1-node (Table 1 without FK) with the referenced node Table 2.
This can be seen in Figure \ref{fig:CompareRM} (a).

A join-table in the RM is existence dependent on the tables it refers to by FKs. 
The FKs forming the primary key (PK) of the join table are not necessary in the TGM because of the same reason as mentioned above.

In Figure \ref{fig:CompareRM} (b) the join-table RST maps directly to an hyper-edge labeled RST with property $col_4$. 
The hyper-edge and the join-table in the relational case represent a many-to-many connection linking 3 
nodes that correspond to the respective tables in the RM. 
In the relational model it is not possible to restrict this ternary relationship to (*,1,1). 
For instance, if 
many connections of Table 1 should link to exactly \emph{one} connection of Table 2 and \emph{one} connection of Table 3. For the TGM this would be simple; only the cardinality at Table 2 and Table 3 would have to be changed from 1..* to 1.

To make the above ternary relationship example less abstract the RST could be an offer of products from Table 1 from one supplier of Table 3 to the client of Table 2. 
With this in mind it is clear that an offer depends on the product(s), one supplier, and one client.

The TGM can also represent non-normalized tables because the model supports complex structured data types having multivalued or array data. 
It is only necessary to define the appropriate data types in the set of available data types $T$.

\subsection{Comparison with XML Schema}
\label{ssec:XML}

XML documents represent hierarchical hypertext documents. 
The document structure can be defined by an XML schema. The hierarchy of XML-documents is directly supported by the TGM using directed edges. XLink provides references (arcs) between elements of internal or external XML-documents. Extended XLinks can connect to more than one element, but the references are always instance based, i. e. the target elements must be listed by URI. 
The TGM is more abstract and expressive allowing the definition of non-hierarchical references on the schema level.

As example for the comparison serves a bookstore offering an unlimited number of books.
A simple XML-schema for the bookstore is given by w3schools.com.
The schema defines books with elements like "title", "author", etc. and its corresponding data types. 
Some data types are not as precise as they could, e. g. the data type xs:double for the price element. We will replace xs:double in our TGM by the money-type \emph{euro} to be more precise.
Some elements have attributes attached like the language ("lang") of a book title. 
The attribute minOccurs="1" of xs:sequence requires the bookstore to have a least one book.

\begin{footnotesize}
\begin{verbatim}
<?xml version="1.0" encoding="utf-8"?>
<xs:schema ... >
  <xs:element name="bookstore" >
    <xs:complexType >
      <xs:sequence minOccurs="1" 
          maxOccurs ="unbounded" >
        <xs:element name="book" >
          <xs:complexType>
            <xs:sequence>
              <xs:element name="title" >
              <xs:complexType>
                <xs:simpleContent>
                  <xs:extension base="xs:string">
                    <xs:attribute name="lang"
                        type="xs:string" />
                  </xs:extension>
                </xs:simpleContent>
              </xs:complexType>
                </xs:element>
              <xs:element name="author" 
                  type ="xs:string"/>
              <xs:element name="year" 
                  type ="xs:integer"/>
              <xs:element name="price" 
                  type ="xs:double"/>
            </xs:sequence>
            <xs:attribute name="category"
                type="xs:string"/>
          </xs:complexType>
        </xs:element>
      </xs:sequence>
    </xs:complexType>
  </xs:element>
</xs:schema>
\end{verbatim}
\end{footnotesize}

If we model the XML-elements as nodes in TGM then XML-attributes and the element values should be represented as properties. 
The name of an XML-element is mapped to a node label. The order of the XML-elements cannot be represented with this approach and XML-element values can be distinguished from XML-attributes by convention only. 

An alternative TGM model represents the complete book structure as one node. 
In this case the XML-elements and their attributes are modeled as structured properties of the book. 
The order of the elements and their associated attributes can be preserved. 
In fact, if XML Schema is used for specifying the data types $N_S$ and $E_S$ (see Subsection \ref{ssec:Schema}) all the flexibility and semantics provided by XML Schema can be represented with the TGS. This argument shows that the TGM is at least as powerful as the XML model.

\begin{figure}[]
\centering
\includegraphics[width=0.48\textwidth]{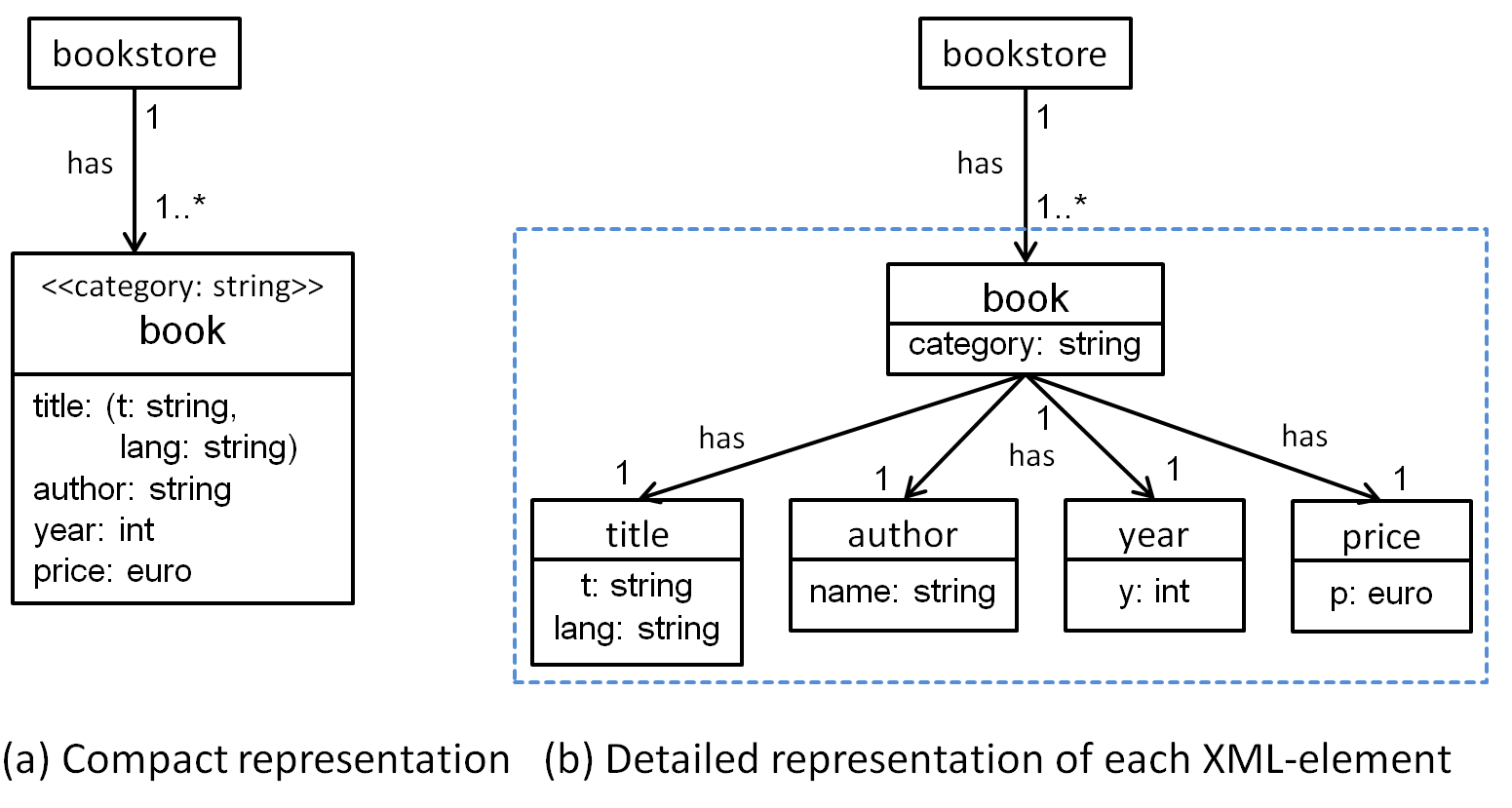}
\caption{Comparing an XML Schema example with TGM}
\label{fig:CompareXML}
\end{figure}

The example bookstore is depicted in Figure \ref{fig:CompareXML} where the left part (a) shows the compact version with the whole book modeled as one node and the right part (b) shows the version where each XML-element is modeled as node. 
We see from this example another possibility to use sub-graphs for higher abstracted models. 

\subsection{Comparison with the Object-Oriented Model}
\label{ssec:OOM}

Because we already use the UML for rendering the TGM, it is easy to see that classes correspond one-to-one with typed hyper-nodes. 
Any methods are simply ignored as we only deal with the network structure of OOM. 
Any complex internal class structure can be directly modeled by appropriate data types $t \in T$. 
The type set $T$ is defined beforehand but can contain any user defined structures. 
In contrast to the OOM the TGM allows different levels of abstraction in the modeling depending whether a structure is modeled by a detailed graph with simple types or a more compact graph using complex data types. This shows the same semantic expressiveness for structures, but a higher flexibility of the TGM. Considering the operations on data the OOM has the advantage to specify the allowed operations by methods.

The UML provides a rich set of association types, which need to be mapped to the label of the edges. 
Our TGM provides types not only for nodes but also for edges (called associations in UML). 
With this information it is possible to model different association types like aggregation, generalization, etc. 
Even user defined associations are possible, e. g., an aggregate could be further qualified as un-detachable or detachable composition or a loose containment.  
The arrow of the edge only indicates the reading direction of the association but does not limit the navigation of the TGM.

\begin{figure}[]
\centering
\includegraphics[width=0.48\textwidth]{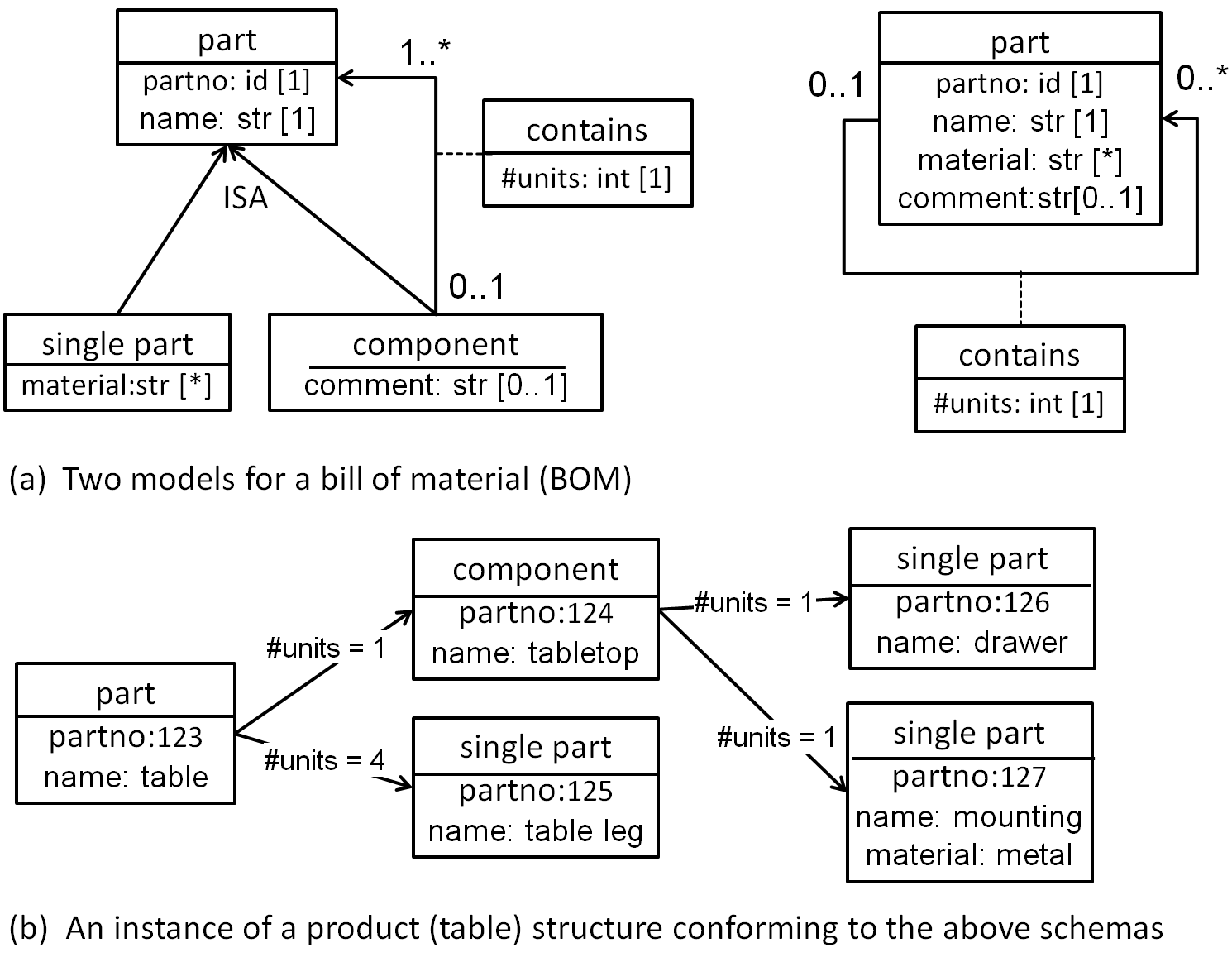}
\caption{Comparison by example with the OOM}
\label{fig:CompareOOM}
\end{figure}

It is also possible to model recursive structures as the examples from Figure \ref{fig:CompareOOM} illustrates. 
The bill of material (BOM) is an important example for a recursive structure used in production planning and control. 
It defines recursively a (compound) part with its components until a single part is reached. 
As example instance a \emph{table} is given in Figure \ref{fig:CompareOOM} (b) consisting of 4 table legs and a tabletop consisting of a drawer and a mounting.

If the edge of \emph{contains} in Figure \ref{fig:CompareOOM} (a) is followed against the arrow direction it is possible to find the component where an individual part is built-in (used).
A complete \textit{where-used list} for a generic (not an individual) part may be obtained with a small schema modification. 
Only the from-end of the \emph{contains}-edge needs to change its multiplicity from 0..1 to 0..*. 
With this small modification all components can be identified where a generic part is used. 

\subsection{Comparison with RDF Schema}
\label{ssec:RDFS}

RDF is a data format for expressing statements about resources with emphasis on Web data. 
A \emph{resource} can be anything, including data, objects (incl. people), and (abstract) concepts.   
RDF Schema uses a semantic extension (meta-constructs) of the basic RDF vocabulary to model RDF data. 
Simple statements about a resource are expressed in the form of \emph{subject-predicate-object} triples. 
It is clear that the underlying structure of such RDF-triples is graph based where a node represents a \emph{subject} or an \emph{object} and an edge represents the \emph{predicate}. 
This implies that RDF Schema triples can be expressed as property graph.

\begin{figure}[]
\centering
\includegraphics[width=0.35\textwidth]{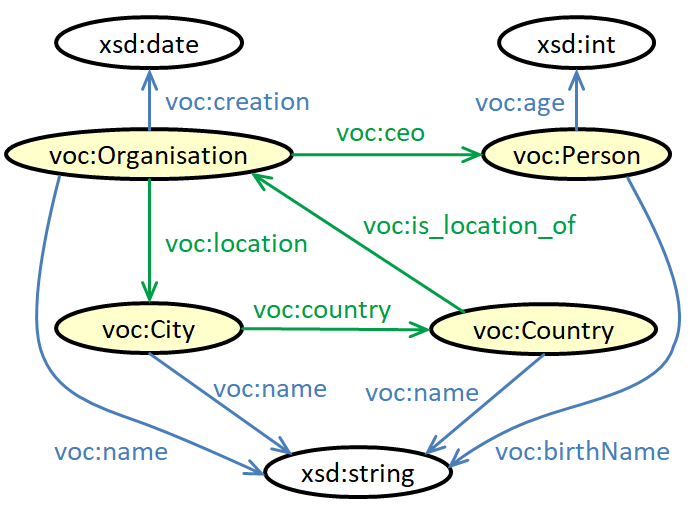}
\caption{Graphical Illustration of the Example RDFS from Angles \cite{Angles2020} }
\label{fig:RDFSchema}
\end{figure}

\begin{figure}[]
\centering
\includegraphics[width=0.45\textwidth]{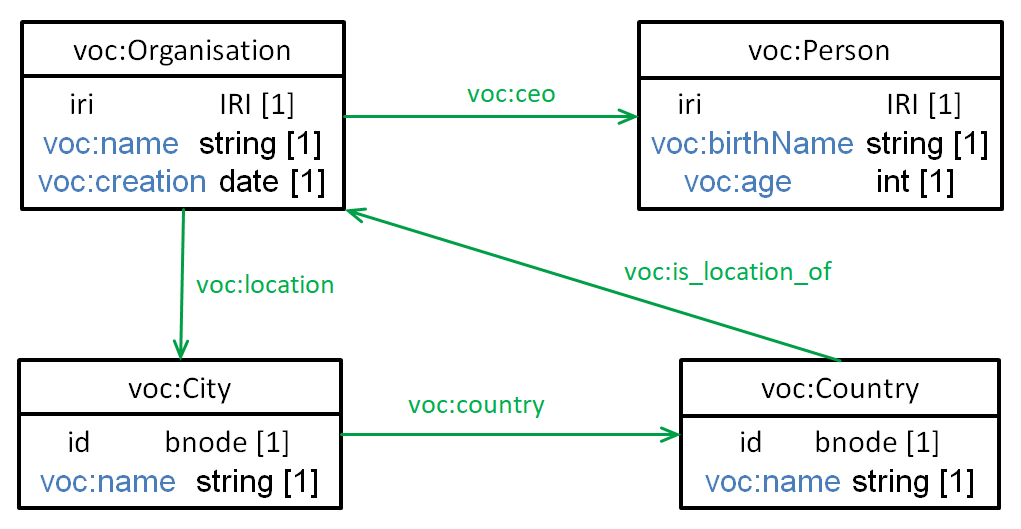}
\caption{Resulting property graph schema of example Figure \ref{fig:RDFSchema}  }
\label{fig:PGSchema}
\end{figure}

Angles et al. \cite{Angles2020} describe a schema-dependent computable schema mapping from an RDF Schema to a Property Graph Schema (PGS) that is semantics and information preserving. 
They prove in their paper that "the PGM subsumes the information capacity of the RDF data model".
The mapping is illustrated in their paper with an example RDF Schema depicted in Figure \ref{fig:RDFSchema}. 
Because of the asymmetrical definition of subject (may not be a literal) and object (can be a literal) the mapping distinguishes objects that are also subjects and objects that are "only objects".
The yellow colored subjects are mapped to nodes in the PGS and the "only objects" (the data types xsd:date, xsd:int, and xsd:string in their example) are mapped to PGS properties.
The green colored predicates are RDF predicates which are mapped to PGS edges, whereas the blue colored predicates are properties of the subject with a data type specified by an "only object".
The resulting property graph schema has 4 nodes and 4 edges and is visualized in Figure \ref{fig:PGSchema} as TGS using the same coloring scheme as in the RDF schema. 

When comparing the RDF Schema with the TGS we see that both share most characteristics. 
In particular, nodes and edges have unique labels. 
These labels serve as global identifiers, called International Resource Identifier (IRI) in the RDF syntax. 


\begin{figure*}[]
\centering
\includegraphics[width=0.95\textwidth]{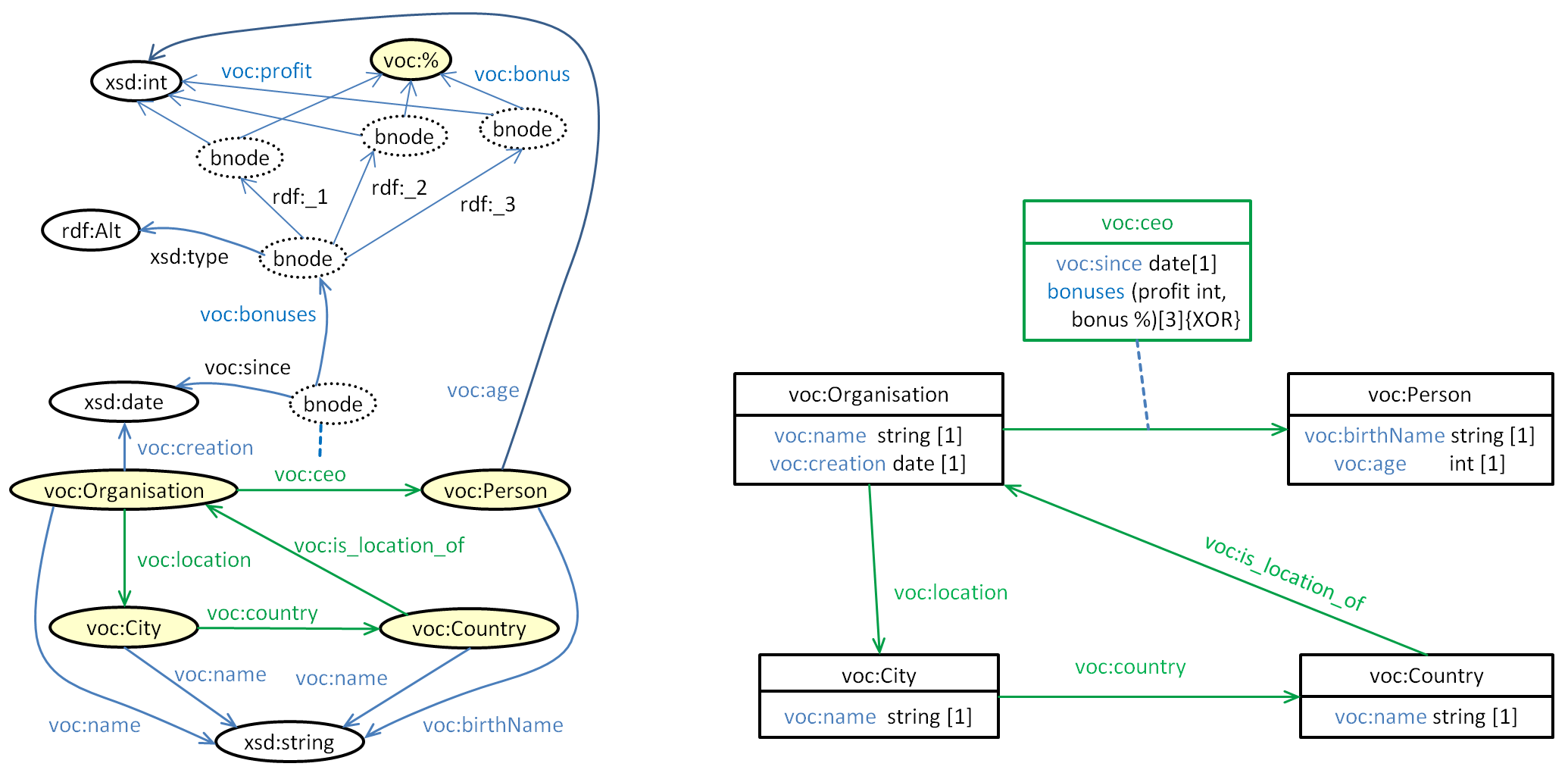}
\caption{Example RDFS (left part) to TGS (right part) mapping with complex predicates}
\label{fig:RDFtoTGM}
\end{figure*}

There are some limitations of the PGM that the TGM overcomes:
\begin{itemize}
\item Properties of RDF graphs support multi-value properties, whereas the PGM usually supports only single value properties \cite{Angles2020}. 
The TGM, however, uses the appropriate data-type to model multi-value properties. 
This can be array, set, list, or bag data types.     
\item In RDF edges can have edges, i. e. edges hold meta-data. This is realized with (edge, predicate, object) triples. 
The simplest situation for this is if a predicate has a label. 
The TGM handles this and more complex situations by edge properties (see Figure \ref{fig:RDFtoTGM} for an example) or hyper-edges, which is graphically rendered by an UML association class. 
\item RDF has three types of nodes (IRI, blank node, and literal) which need to be mapped to only one node type in the PGM. 
Because the TGM supports typed nodes any RDF node-type can be handled directly. 
\item The RDF model has some special semantics like reification and subclassing (rdfs:subClassOf and rdfs:subPropertyOf) that are not supported in the paper of Angles et al. \cite{Angles2020}.
The TGM also supports these special structures. This is achieved with the appropriate node or edge types.
\item Each node or edge in an RDF graph contains one single IRI, a Literal or nothing (in case of a blank node), whereas each node or edge in a TGM could contain multiple properties, depending on its type.
This includes multi-value properties and properties for hyper-edges as explained before.
\end{itemize}

With the TGM any kind of node type can be supported because of its capability to define user defined types for the nodes and edges.
RDF allows to identify an RDF triple (statement) with an IRI. Such an aggregate construct is possible in TGM with a ternary edge type.

For a general schema mapping we have the following rules:
\begin{enumerate}
\item Abstract elements (subjects) are mapped to TGM nodes.
\item Aggregation (predicates) and generalization (rdfs:subClassOf, rdfs:subPropertyOf) elements are mapped to TGM edges with the appropriate type.
\item Functions (predicate cardinality is always 1) are mapped to TGM edge ends. 
\item Lexical elements (literals, IRIs, "only objects") are mapped to TGM properties.
\end{enumerate}

In order to illustrate the mapping to TGM we amend the example from Angles et al. \cite{Angles2020} with more properties for the voc:ceo predicate.
The predicate now holds the date when the CEO was appointed and the profit dependent bonus payments.
For each voc:profit predicate a voc:bonus is associated. The profit is of type xsd:int and the bonus has a user defined percent data type with a range from 0 to 100. 
RDF auxiliary nodes (bnodes) are needed to add this information pair without any ambiguity. 
We need a bnode of type rdf:Alt to indicate that the bonuses are alternatives. 

Figure \ref{fig:RDFtoTGM} shows on the left side the amended RDFS example and the corresponding TGS on the right side. 
It immediately catches the eye that the TGS is far more readable for humans than the RDFS, which make it better suitable for data modeling.
It is cumbersome to add properties to a predicate as it requires blank nodes (bnodes) in order to form legal and unambiguous RDF triples. 
For the corresponding TGM edge property we simply use an array of size 3 with a structured data type containing profit and the corresponding bonus in percent. 
The array \emph{bonuses} can have a constraint XOR that clarifies the semantics as an alternative 1 out of 3.
The comparison of both schemata clearly shows that even with the coloring the RDFS tends to quickly becomes confusing because properties are not aggregated to objects or predicates. 
Also data types are expressed on the same level as subjects, which makes the RDF less clear for data modeling purposes.
  
\section{Conclusion and Future Work}
In this paper we presented a structure definition of the TGM and an UML-like notation to visualize a graph database and its graph schema. 
Due to the TGS with predefined and user-defined data types, the TGM improves the formal data quality compared to other graph models.
We have demonstrated the superior modeling power in comparison to other graph data models and prevalent data models, namely relational, object oriented, XML model, and RDFS.
The model supports built in and user defined complex data types, which allow different abstraction levels. 
Another possibility for abstraction is to compress a sub-graph into a hyper-node reducing the visible complexity.
This capability is especially useful for large and complex data models.
   
Because of its semantic modeling power the TGM could act as \textbf{supermodel} for model-management and serve as a unifying data model that supports data integration from various data sources with different data models. 
The main challenge for an automated data integration are incompatible data sources where the TGM as \textbf{supermodel} could help to solve quality issues and specify information preserving data translations.
Details, like a score for the mapping quality and how much of the information was preserved by the mapping still need to be investigated.
The development of a manipulation and query language for the TGM is future work. 
The idea is to combine elements of other graph languages with the dot-notation known from object-oriented languages to support navigation paths.

\end{document}